# Multidimensional Realization Theory and Polynomial System Solving

Philippe Dreesen[*,a], Kim Batselier[b], and Bart De Moor[c,d]

[a]Vrije Universiteit Brussel (VUB), Dept. VUB-ELEC, Brussels, Belgium
[b]The University of Hong Kong, Dept. Electrical and Electronic Engineering, Hong Kong
[c]KU Leuven, Dept. Electrical Engineering (ESAT), STADIUS Center for Dynamical Systems, Signal Processing and Data Analytics, Leuven, Belgium
[d]imec, Leuven, Belgium

## Abstract

Multidimensional systems are becoming increasingly important as they provide a promising tool for estimation, simulation and control, while going beyond the traditional setting of one-dimensional systems. The analysis of multidimensional systems is linked to multivariate polynomials, and is therefore more difficult than the well-known analysis of one-dimensional systems, which is linked to univariate polynomials. In the current paper we relate the realization theory for overdetermined autonomous multidimensional systems to the problem of solving a system of polynomial equations. We show that basic notions of linear algebra suffice to analyze and solve the problem. The difference equations are associated with a Macaulay matrix formulation, and it is shown that the null space of the Macaulay matrix is a multidimensional observability matrix. Application of the classical shift trick from realization theory allows for the computation of the corresponding system matrices in a multidimensional state-space setting. This reduces the task of solving a system of polynomial equations to computing an eigenvalue decomposition. We study the occurrence of multiple solutions, as well as the existence and analysis of solutions at infinity, which allow for an interpretation in terms of multidimensional descriptor systems.

## 1 Introduction

Recent years have witnessed a surge in research on multidimensional systems theory, identification and control (Batselier & Wong, 2016; Bose, 2007; Hanzon & Hazewinkel, 2006a; Ramos & Mercère, 2016; Rogers et al., 2015; Zerz, 2000, 2008). There is a broad scientific interest regarding multidimensional systems, as they offer an extension to the well-known class of one-dimensional linear systems, in which the system trajectories depend on a single variable (such as time or frequency), to a dependence on several independent variables (such as a two-dimensional position, spatio-temporal systems, parameter varying systems, etc.). However, the analysis of multidimensional systems is known to be more complicated than that of one-dimensional systems.

For one-dimensional systems it is well-known that the Laplace transform or the Z-transform (Kailath, 1980) can be used to relate with the system description a polynomial formulation. This connection is central in systems theory and its applications. For multidimensional systems, the analysis is more difficult as it involves multivariate polynomials, and hence the tools of (computational) algebraic geometry or differential algebra (Buchberger, 2001; Hanzon & Hazewinkel, 2006a). Nevertheless, several multidimensional models and their properties have been studied extensively (Attasi, 1976; Bose, 1982; Bose, Buchberger, & Guiver, 2003; Fornasini, Rocha, & Zampieri, 1993; Gałkowski, 2001; Kaczorek, 1988; Kurek, 1985; Livšic, 1983; Livšic, Kravitsky, Markus, & Vinnikov, 1995; Oberst, 1990; Roesser, 1975), and applications in identification (Ramos & Mercère, 2016) and control (Rogers et al., 2015) are known.

---

[*]Corresponding author. Email: philippe.dreesen@gmail.com

The current article studies a specific class of multidimensional systems, namely *overdetermined multidimensional systems* (Ball, Boquet, & Vinnikov, 2012; Ball & Vinnikov, 2003; Batselier & Wong, 2016; Fornasini et al., 1993; Hanzon & Hazewinkel, 2006b; Rocha & Willems, 2006; Shaul & Vinnikov, 2009), and aims at exposing some interesting, yet largely unknown links with linear algebra and polynomial system solving. Specifically, we relate realization theory for discrete-time overdetermined autonomous systems to the task of solving a system of polynomial equations.

Overdetermined multidimensional systems have the restriction that there are compatibility constraints on the input and output signals (Ball & Vinnikov, 2003), e.g., for autonomous systems this compatibility condition is expressed in the fact that the system matrices of its state-space formulation must commute. Overdetermined systems were originally studied in a continuous-time framework (Ball & Vinnikov, 2003; Livšic, 1983; Livšic et al., 1995), but also recently in a discrete-time framework (Batselier & Wong, 2016; Bleylevens, Peeters, & Hanzon, 2007; Dreesen, 2013; Hanzon & Hazewinkel, 2006b). In the current paper, we will study discrete-time autonomous overdetermined systems, which are given in a state-space formulation as

$$
\begin{array}{rcl}
\boldsymbol{x}[k_1+1, k_2, \ldots, k_n] & = & \boldsymbol{A}_1 \boldsymbol{x}[k_1, \ldots, k_n] \\
& \vdots & \\
\boldsymbol{x}[k_1, \ldots, k_{n-1}, k_n+1] & = & \boldsymbol{A}_n \boldsymbol{x}[k_1, \ldots, k_n] \\
y[k_1, \ldots, k_n] & = & \boldsymbol{c}^\top \boldsymbol{x}[k_1, \ldots, k_n]
\end{array}
\tag{1}
$$

where $\boldsymbol{x} \in \mathbb{R}^m$ is an $m$-dimensional state vector that depends on $n$ independent indices, the matrices $\boldsymbol{A}_i \in \mathbb{R}^{m \times m}$ define the autonomous state transitions, and $\boldsymbol{c} \in \mathbb{R}^m$ defines how the one-dimensional output $y$ is composed from the state vector $\boldsymbol{x}$.

It is important to remark that the class of overdetermined multidimensional systems (1) is rather different than the more commonly used multidimensional systems of Roesser (1975) and Fornasini and Marchesini (1976). For instance, in the Roesser model, the state vector $\boldsymbol{x}$ is divided into partial state vectors along each 'direction', which is not the case in overdetermined systems. Also, both the Roesser and Fornasini-Marchesini models require an infinite number of initial states in order to compute the state recursion, which is not the case in overdetermined systems (Batselier & Wong, 2016).

The central question that is tackled in the current paper, is *how a state-space realization can be obtained from a given set of n difference equations.* This problem formulation is in the same spirit as the classical multidimensional realization problem, where from a given transfer function description, a state-space representation is sought (Gałkowski, 2001; Xu, Fan, Lin, & Bose, 2008; Xu, Yan, Lin, & Matsushita, 2012).

We will explore the realization problem from a linear algebra point-of-view and will show how a natural link emerges between applying realization theory inspired by the algorithm of Ho and Kalman (1966) and the Macaulay resultant-based matrix method for polynomial system solving (Cox, Little, & O'Shea, 2005; Jónsson & Vavasis, 2004; Macaulay, 1916; Mourrain, 1998). We will show that the Macaulay matrix formulation contains (multidimensional) time-shifted difference equations. We will then highlight natural and accessible links between multivariate polynomials and multidimensional realization theory. In particular we will illustrate that:

- admissible output trajectories are elements of the null space of the Macaulay matrix;

- the null space of the Macaulay matrix is a multidimensional observability matrix;

- applying the shift trick of realization theory yields a state-space realization of the system;

- the state-space realization of the system corresponds to the Stetter eigenvalue formulation;

- roots with multiplicities give rise to partial derivative operators in the null space;

- solutions at infinity can be analyzed by phrasing the problem in projective space; and

- solutions at infinity can be related to a descriptor system realization.

It is known that a system of multivariate polynomial equations and its corresponding Groebner basis have a straightforward interpretation as a state-space realization of the associated difference equations (Fornasini et al., 1993; Hanzon & Hazewinkel, 2006b). From this observation, the current article will employ develop a resultant matrix-based link with realization theory that results an eigenvalue-based root-finding method. Notice that this is a system-theoretical interpretation of Stetter's eigenvector method, which has been discovered independently by several researchers in the 1980s and 1990s (Auzinger & Stetter, 1988; Lazard, 1983; Möller & Stetter, 1995; Mourrain, 1998; Stetter, 2004). Wheras Fornasini et al. (1993); Hanzon and Hazewinkel (2006b) employ a Groebner basis approach to find the system matrices, the proposed method in this article uses a linear algebra formulation and does not require the computation of a Groebner basis, and is therefore more reminiscent of matrix-based methods like the ones of Jónsson and Vavasis (2004), and Mourrain (1998), among others. Furthermore, we will discuss the occurrence of solutions at infinity, and their system theoretical interpretation involving a descriptor system realization.

The remainder of this article is organized as follows. In Section 2 the links between one-dimensional systems and polynomial root-finding are reviewed, employing the Sylvester matrix formulation, providing a blueprint to generalize the matrix approach to the multivariate case. In Section 3 the Macaulay matrix formulation is introduced and given an interpretation in the context of multidimensional systems. Section 4 illustrates how the well-known shift trick from realization theory can be applied to the null space of the Macaulay matrix to obtain a multidimensional realization. This is equivalent to phrasing the Stetter eigenvalue problem for polynomial root-finding. In Section 5 it is shown how solutions at infinity can be separated from affine solutions. This separation is given a system theoretic interpretation as a splitting into a regular and a descriptor system. In Section 6 we draw the conclusions of this work and point out problems for further research.

We have aimed to keep the exposition as simple and accessible as possible, requiring only the most elementary notions of linear algebra and state-space system theory. Throughout the paper, systems of polynomial equations are used to represent multidimensional difference equations (using a multi-indexed Z-transform). We assume that the difference equations define scalar signals that vary in $n$ independent indices. Furthermore, we assume that the corresponding systems of polynomials describe zero-dimensional solution sets in the projective space, a fact that our approach will also reveal as the dimensions of certain null spaces will stabilize in the case the solution set is zero-dimensional. With some abuse of terminology, at times we may refer to an overdetermined system of multidimensional difference equations as its representation as a polynomial system, or vice versa.

## 2 One-dimensional systems lead to univariate polynomials

In the current section we will review 'by example' a few well-known facts from linear algebra and system theory that tie together polynomial root-finding and realization theory. These examples serve to introduce the tools we will use in the remainder of the paper. We may switch back and forth between the polynomial system solving and the multidimensional systems settings depending on which one is more natural for a specific aspect of our exposition.

**Example 1.** By introducing the shift operator $z$ that is defined as $(zw)[k] = w[k+1]$, one can associate with the difference equation $w[k+2] - 3w[k+1] + 2w[k] = 0$ the polynomial equation $p(z) = z^2 - 3z + 2 = 0$. We write $p(z) = 0$ as its vector of coefficients multiplied by a Vandermonde monomial vector $\boldsymbol{v}$ as $\boldsymbol{p}^\top \boldsymbol{v} = \begin{bmatrix} 2 & -3 & 1 \end{bmatrix} \begin{bmatrix} 1 & z & z^2 \end{bmatrix}^\top = 0$. In terms of the difference equation, this expression is nothing more than $\begin{bmatrix} 2 & -3 & 1 \end{bmatrix} \begin{bmatrix} w[k] & w[k+1] & w[k+2] \end{bmatrix}^\top = 0$

The roots of $p(z)$ are $z^{(1)} = 1$ and $z^{(2)} = 2$, and they can be computed by means of linear algebra as follows: The two solutions generate two vectors that span the right null space of $\boldsymbol{p}^\top$, which we call the Vandermonde basis $\boldsymbol{V}$ of the null space. We have $\boldsymbol{p}^\top \boldsymbol{V} = \boldsymbol{0}^\top$ with

$$\boldsymbol{V} = \begin{bmatrix} 1 & 1 \\ z^{(1)} & z^{(2)} \\ (z^{(1)})^2 & (z^{(2)})^2 \end{bmatrix} = \begin{bmatrix} 1 & 1 \\ 1 & 2 \\ 1 & 4 \end{bmatrix}. \tag{2}$$

The Vandermonde basis $V$ has a multiplicative shift structure, allowing us to write

$$\underline{V}D = \overline{V}, \tag{3}$$

where $D = \text{diag}(z^{(1)}, z^{(2)})$ and $\overline{V}$ and $\underline{V}$ denotes $V$ with its first and last row removed, respectively. This is a direct application of the shift trick of realization theory in the null space of $p^\top$ (Ho & Kalman, 1966; Willems, 1986b).

In practice, the Vandermonde basis $V$ cannot be obtained directly, but instead any (numerical) basis for the null space can be used. Indeed, the shift structure is a property of the column space of $V$, and is hence independent of the choice of basis. Thus, the shift relation holds for any basis $Z$ of the null space, which is related to $V$ by a nonsingular transformation $T$ as $V = ZT$, leading to the generalized eigenvalue equation

$$\underline{Z}TD = \overline{Z}T. \tag{4}$$

Another choice of basis that is worth mentioning is obtained by putting a numerical basis $Z$ of the null space in its column echelon form $H$. Therefore, let us first recall how a numerical basis of the null space of a matrix $M$ is found using the singular value decomposition.

**Lemma 1** (Numerical basis of the null space). *A numerical basis of the null space $Z$ can be obtained from the singular value decomposition from*

$$M = \begin{bmatrix} U_1 & U_2 \end{bmatrix} \begin{bmatrix} \Sigma & \\ & 0 \end{bmatrix} \begin{bmatrix} W^\top \\ Z^\top \end{bmatrix}.$$

The column echelon basis of the null space $H$ can be constructed in a classical 'Gaussian elimination' fashion, or by means of numerical linear algebra as follows.

**Lemma 2** (Column echelon form). *Let $Z$ be a numerical basis of the null space of $M$, e.g., computed with the singular value decomposition. Let $Z^\star$ be composed of the linearly independent rows of $Z$, where linear independence is checked going from the top to the bottom rows, ordered by the degree negative lexicographic order. The column reduced echelon form is given by $H = Z(Z^\star)^\dagger$. Remark that computing $H$ may be numerically ill-posed as it requires checking linear (in)dependence of single rows of $Z$.*

An important property is that the column echelon form is related to $V$ by the relation $V = HU$, with

$$\begin{bmatrix} 1 & 1 \\ 1 & 2 \\ 1 & 4 \end{bmatrix} = \begin{bmatrix} 1 & 0 \\ 0 & 1 \\ -2 & 3 \end{bmatrix} \begin{bmatrix} 1 & 1 \\ 1 & 2 \end{bmatrix}, \tag{5}$$

where we notice that the columns of $U$ have the form $\begin{bmatrix} 1 & z \end{bmatrix}^\top$, evaluated in the solutions $z^{(1)} = 1$ and $z^{(2)} = 2$ (this fact will be proven later on in Proposition 1). Applying the shift relation in the column echelon basis $H$ leads to the well-known Frobenius companion matrix formulation

$$\underline{H}UD = \overline{H}U \Leftrightarrow \begin{bmatrix} 1 & 0 \\ 0 & 1 \end{bmatrix} UD = \begin{bmatrix} 0 & 1 \\ -2 & 3 \end{bmatrix} U. \tag{6}$$

Remark that, applying the shift relation on the column echelon basis $H$ of the null space results in the well-known Frobenius companion matrix form.

It is important to remark that a basis of the null space (in fact, *any* basis of the null space) can be identified with an (extended) observability matrix $\mathcal{O}_2$ of the system described by the difference equation, where

$$\mathcal{O}_2 = \begin{bmatrix} c^\top \\ c^\top A \\ c^\top A^2 \end{bmatrix}, \tag{7}$$

where the corresponding state-space model is given as

$$\begin{aligned} x[k+1] &= Ax[k], \\ y[k] &= c^\top x[k]. \end{aligned} \tag{8}$$

From this we can extract an associated state-space realization by letting $\boldsymbol{c}^\top$ correspond to the first row of $\boldsymbol{\mathcal{O}}_2$, and $\boldsymbol{A}$ is found from $\underline{\boldsymbol{\mathcal{O}}}_2 \boldsymbol{A} = \overline{\boldsymbol{\mathcal{O}}}_2$. We find the state-space description (where we have used $\boldsymbol{H}$ as the observability matrix)

$$\begin{bmatrix} w[k+1] \\ w[k+2] \end{bmatrix} = \begin{bmatrix} 0 & 1 \\ -2 & 3 \end{bmatrix} \begin{bmatrix} w[k] \\ w[k+1] \end{bmatrix},$$

$$y[k] = \begin{bmatrix} 1 & 0 \end{bmatrix} \begin{bmatrix} w[k] \\ w[k+1] \end{bmatrix}. \tag{9}$$

Notice that for the column echelon basis $\boldsymbol{H}$ of the null space the eigenvectors in $\boldsymbol{U}$ are monomial vectors, which allows for a direct interpretation as a state-space realization. In system theoretic terms, the Frobenius matrix chains together the consecutive samples of the trajectory $w$. △

The same procedure can be used to study the solutions of a set of difference equations. In terms of polynomial algebra, this turns out to be equivalent to finding the greatest common divisor of a set of univariate polynomials and can be solved by means of the Sylvester matrix construction. We will illustrate this in the following example.

**Example 2.** Consider two difference equations

$$\begin{aligned} w[k+3] + 2w[k+2] - 5w[k+1] - 6w[k] &= 0, \\ w[k+2] - w[k+1] - 2w[k] &= 0, \end{aligned} \tag{10}$$

which can be associated with the polynomial equations

$$\begin{aligned} p(z) &= z^3 + 2z^2 - 5z - 6 = 0, \\ q(z) &= z^2 - z - 2 \phantom{+ 2z^2 - 5z} = 0. \end{aligned} \tag{11}$$

The common roots of $p(z)$ and $q(z)$ are $z^{(1)} = -1$ and $z^{(2)} = 2$. Finding the $w[k]$ that satisfy both equations quickly leads to the Sylvester matrix construction

$$\begin{bmatrix} -6 & -5 & 2 & 1 & 0 \\ 0 & -6 & -5 & 2 & 1 \\ -2 & -1 & 1 & 0 & 0 \\ 0 & -2 & -1 & 1 & 0 \\ 0 & 0 & -2 & -1 & 1 \end{bmatrix} \begin{bmatrix} 1 \\ z \\ z^2 \\ z^3 \\ z^4 \end{bmatrix} = \begin{bmatrix} 0 \\ 0 \\ 0 \\ 0 \\ 0 \end{bmatrix}, \tag{12}$$

which is obtained by multiplying $p$ and $q$ by powers of $z$. In computer algebra, this problem is known as the greatest common divisor (GCD) problem. Notice that the common roots of $p(z)$ and $q(z)$ give rise to Vandermonde-structured vectors in the null space. Again we have arrived at a point where the solution to the problem involves a Vandermonde structured matrix.

From the system theoretic point of view, the Sylvester matrix construction generates additional equations that impose constraints on $w[k]$, simply by including shifted instances of the given equations until a square system of linear equations is obtained. Remark that *any* vector in the null space of the Sylvester matrix defines a valid trajectory $w[k]$ that satisfies *both* difference equations. Again, a basis for the null space can be associated with an observability matrix $\boldsymbol{\mathcal{O}}$, on which applying the shift trick reveals a state-space realization.

The column echelon basis $\boldsymbol{H}$ of the null space is in this case

$$\boldsymbol{H} = \begin{bmatrix} 1 & 0 \\ 0 & 1 \\ 2 & 1 \\ 2 & 3 \\ 6 & 5 \end{bmatrix}, \tag{13}$$

and applying the shift trick leads to a rectangular generalized eigenvalue problem: We find $\underline{\boldsymbol{H}}\boldsymbol{U}\boldsymbol{D} = \overline{\boldsymbol{H}}\boldsymbol{U}$ as

$$\begin{bmatrix} 1 & 0 \\ 0 & 1 \\ 2 & 1 \\ 2 & 3 \end{bmatrix} \begin{bmatrix} 1 & 1 \\ -1 & 2 \end{bmatrix} \begin{bmatrix} -1 & 0 \\ 0 & 2 \end{bmatrix} = \begin{bmatrix} 0 & 1 \\ 2 & 1 \\ 2 & 3 \\ 6 & 5 \end{bmatrix} \begin{bmatrix} 1 & 1 \\ -1 & 2 \end{bmatrix}. \tag{14}$$

It suffices to reduce the above relation to the square eigenvalue problem from which $\boldsymbol{U}$ and $\boldsymbol{H}$ can be obtained as well. A square generalized eigenvalue problem can be obtained by selecting the first linear independent rows of $\boldsymbol{H}$ as to obtain a square invertible matrix. In this case, the first two rows of $\boldsymbol{H}$ are linearly independent, and hence we find

$$\begin{bmatrix} 1 & 0 \\ 0 & 1 \end{bmatrix} \begin{bmatrix} 1 & 1 \\ -1 & 2 \end{bmatrix} \begin{bmatrix} -1 & 0 \\ 0 & 2 \end{bmatrix} = \begin{bmatrix} 0 & 1 \\ 2 & 1 \end{bmatrix} \begin{bmatrix} 1 & 1 \\ -1 & 2 \end{bmatrix}. \tag{15}$$

We observe that again a Frobenius companion matrix shows up: it can be verified that in this case it is the companion matrix of the GCD of $p(z)$ and $q(z)$. A state-space realization of the common trajectories is given by

$$\begin{bmatrix} w[k+1] \\ w[k+2] \end{bmatrix} = \begin{bmatrix} 0 & 1 \\ 2 & 1 \end{bmatrix} \begin{bmatrix} w[k] \\ w[k+1] \end{bmatrix},$$

$$y[k] = \begin{bmatrix} 1 & 0 \end{bmatrix} \begin{bmatrix} w[k] \\ w[k+1] \end{bmatrix}. \tag{16}$$

△

Although these examples have trivial results, they illustrate the fact that linear algebra and realization theory are natural tools for deriving both eigenvalue-based root-finding methods as well as state-space realizations. Moreover, the analysis of the root-finding and realization theory problems turns out to be very similar. In the following sections, we will generalize these ideas to the multivariate case.

## 3 Multidimensional systems lead to multivariate polynomials

We will generalize the results of Section 2 to the multivariate and multidimensional cases. The construction that we will introduce here is a straightforward generalization of the Sylvester matrix formulation of Section 2.

### 3.1 Macaulay's construction

A system of multivariate polynomials defines trajectories $w[k_1, \ldots, k_n] \in \mathbb{R}$, for $k_1, \ldots, k_n \in \mathbb{N}$, of a multidimensional system represented in the representation $\boldsymbol{r}(\boldsymbol{z})w = \boldsymbol{0}$ (Willems, 1986a, 1986b, 1987), where $\boldsymbol{r} \in \mathbb{R}^{n \times 1}$ and $\boldsymbol{z} = (z_1, \ldots, z_n)$ denotes the multidimensional shift operator

$$z_i : (z_i w)[k_1, \ldots, k_i, \ldots, k_n] = w[k_1, \ldots, k_i + 1, \ldots, k_n]. \tag{17}$$

We denote the corresponding system of multivariate polynomial equations as

$$\begin{aligned} f_1(z_1, \ldots, z_n) &= 0, \\ &\vdots \\ f_n(z_1, \ldots, z_n) &= 0, \end{aligned} \tag{18}$$

having total degrees $d_1, \ldots, d_n$. We assume that (18) has a zero-dimensional solution set.

In the one-dimensional case, the Fundamental Theorem of Algebra states that a univariate degree $d$ polynomial $f(x)$ has exactly $d$ roots in the field of complex numbers. When several of these roots coincide, we say that they occur with multiplicity. This happens if $f(x)$ has a horizontal tangent at the position of a multiple root. The multidimensional counterpart of the Fundamental Theorem of Algebra is called Bezout's theorem (see Cox et al. (2005, pp. 97) and Shafarevich (2013, pp. 246)). This theorem states that a set of equations (18) that describes a zero-dimensional solution set has exactly $m = \prod_i d_i$ solutions in the projective space, counted with multiplicity.

The notion of projective space is required here, because it may happen that solutions are 'degenerate', and occur *at infinity*. For instance two parallel lines (both having degree one) are expected to have a single common root: they can be thought as meeting at infinity.

For now we assume that all solutions are simple (i.e., without multiplicity) and there are no solutions at infinity. This is for the moment for didactic purposes, and we will discuss the general case in Section 4.3 and Section 5. Under these assumptions the number of solutions $m$ is given by $m = \prod_i d_i$ (see Cox et al. (2005, pp. 97) and Shafarevich (2013, pp. 246)), which we call the *Bezout number*.

Before we formally study the Macaulay matrix and its properties, let us introduce the main ideas with a simple example.

**Example 3.** Consider the following system of difference equations

$$4w[k_1+2, k_2] - 16w[k_1+1, k_2] + w[k_1, k_2+2] - 2w[k_1, k_2+1] + 13w[k_1, k_2] = 0, \\ 2w[k_1+1, k_2] + w[k_1, k_2+1] - 7w[k_1, k_2] = 0. \tag{19}$$

By shifting the above equations up to indices $k_i + 2$ we find

$$\begin{bmatrix} 13 & -16 & -2 & 4 & 0 & 1 \\ -7 & 2 & 1 & 0 & 0 & 0 \\ 0 & -7 & 0 & 2 & 1 & 0 \\ 0 & 0 & -7 & 0 & 2 & 1 \end{bmatrix} \begin{bmatrix} w[k_1, k_2] \\ w[k_1+1, k_2] \\ w[k_1, k_2+1] \\ w[k_1+2, k_2] \\ w[k_1+1, k_2+1] \\ w[k_1, k_2+2] \end{bmatrix} = \begin{bmatrix} 0 \\ 0 \\ 0 \\ 0 \end{bmatrix}. \tag{20}$$

Correspondingly, we may consider the system of equations

$$\begin{array}{rcl} f_1(z_1, z_2) & = & 4z_1^2 - 16z_1 + z_2^2 - 2z_2 + 13 = 0, \\ f_2(z_1, z_2) & = & 2z_1 + z_2 - 7 = 0. \end{array} \tag{21}$$

It can be verified that the solutions are $\left(z_1^{(1)}, z_2^{(1)}\right) = (3, 1)$ and $\left(z_1^{(2)}, z_2^{(2)}\right) = (2, 3)$. The system is represented as

$$\left[ \begin{array}{c|cc|ccc} 13 & -16 & -2 & 4 & 0 & 1 \\ -7 & 2 & 1 & 0 & 0 & 0 \\ 0 & -7 & 0 & 2 & 1 & 0 \\ 0 & 0 & -7 & 0 & 2 & 1 \end{array} \right] \begin{bmatrix} 1 \\ z_1 \\ z_2 \\ z_1^2 \\ z_1 z_2 \\ z_2^2 \end{bmatrix} = \begin{bmatrix} 0 \\ 0 \\ 0 \\ 0 \end{bmatrix}. \tag{22}$$

The Macaulay matrix has dimensions $4 \times 6$, rank four and nullity two. The Vandermonde basis $\boldsymbol{V}$ of the null space is

$$\boldsymbol{V} = \left[ \begin{array}{c|c} 1 & 1 \\ z_1^{(1)} & z_1^{(2)} \\ z_2^{(1)} & z_2^{(2)} \\ z_1^{(1)} z_1^{(1)} & z_1^{(2)} z_1^{(2)} \\ z_1^{(1)} z_2^{(1)} & z_1^{(2)} z_2^{(2)} \\ z_2^{(1)} z_2^{(1)} & z_2^{(2)} z_2^{(2)} \end{array} \right] = \begin{bmatrix} 1 & 1 \\ 2 & 3 \\ 3 & 1 \\ 4 & 9 \\ 6 & 3 \\ 9 & 1 \end{bmatrix}, \tag{23}$$

where the columns are multivariate Vandermonde monomial vectors evaluated at the two solutions $(2, 3)$ and $(3, 1)$. Returning to the system theoretic interpretation, trajectories $w[k_1, k_2]$ that are compatible with (19) are a linear combination of the two basis vectors in (23). △

We will now formally introduce the Macaulay matrix $\boldsymbol{M}_d$ and the corresponding multivariate Vandermonde monomial vector $\boldsymbol{v}_d$ to represent a system of polynomial equations as a system of homogeneous linear equations.

**Definition 1** (Vandermonde monomial vector). The multivariate Vandermonde monomial vector $\boldsymbol{v}_d$ is defined as

$$\boldsymbol{v}_d := \begin{bmatrix} 1 \mid z_1 & z_2 & \ldots & z_n \mid z_1^2 & z_1 z_2 & z_1 z_3 & \ldots & z_n^2 \mid \ldots \mid z_1^d & \ldots & z_n^d \end{bmatrix}^\top. \quad (24)$$

The polynomial $f_i(z_1, \ldots, z_n)$ can in this way be represented as a row vector containing the coefficients multiplied by a Vandermonde vector of a suitable total degree as $\boldsymbol{f}_i^\top \boldsymbol{v}_d$.

The Macaulay matrix contains as its rows such coefficient vectors that are obtained by multiplying the equations $f_i(z_1, \ldots, z_n)$ by monomials such that at most some predefined total degree $d$ is not exceeded.

**Definition 2** (Macaulay matrix). The Macaulay matrix $\boldsymbol{M}_d$ contains as its rows the vector representations of the shifted equations $\boldsymbol{z}^{\boldsymbol{\alpha}_i} \boldsymbol{f}_i^\top$ as

$$\boldsymbol{M}_d := \begin{bmatrix} \{\boldsymbol{z}^{\boldsymbol{\alpha}_1}\} \boldsymbol{f}_1^\top \\ \vdots \\ \{\boldsymbol{z}^{\boldsymbol{\alpha}_n}\} \boldsymbol{f}_n^\top \end{bmatrix}. \quad (25)$$

where each $f_i$, for $i = 1, \ldots, n$ is multiplied by all monomials $\boldsymbol{z}^{\boldsymbol{\alpha}_i}$ of total degrees $\leq d - d_i$, resulting in the assignment of the coefficients of $f_i$ to a position in $\boldsymbol{M}_d$.

The rows of the Macaulay matrix for total degree $d$ represent polynomial consequences of the polynomials $f_1, \ldots, f_n$ that can be obtained by elementary row operations. Remark that the row span of the Macaulay matrix does *not* necessarily coincide with the elements of the ideal $I = \langle f_1, \ldots, f_n \rangle$ of total degree $d$ or less (denoted $I_{\leq d}$) (Cox, Little, & O'Shea, 2007). It is possible that by reductions of degree $\Delta > d$ polynomials, degree $\delta \leq d$ equations are obtained that cannot be reached by the row space of total degree $\delta \leq d$ shifts of the $f_i$.

For the case there are only affine roots, the Macaulay matrix is constructed for total degree $d = \sum_i d_i - n + 1$ (Giusti & Schost, 1999; Lazard, 1983). Henceforth, the dependence of $\boldsymbol{M}_d$ and $\boldsymbol{v}_d$ on $d$ is often left out for notational convenience, i.e., $\boldsymbol{M} := \boldsymbol{M}_d$ and $\boldsymbol{v} := \boldsymbol{v}_d$. It is important to observe that every solution of (18), denoted $\left(z_1^{(k)}, \ldots, z_n^{(k)}\right)$, for $k = 1, \ldots, m$, gives rise to a Vandermonde vector

$$\begin{bmatrix} 1 \mid z_1^{(k)} & \cdots & z_n^{(k)} \mid \cdots \mid (z_1^{(k)})^d & \cdots & (z_1^{(k)})^d \end{bmatrix}^\top \quad (26)$$

in the null space of $\boldsymbol{M}$. The collection of all such vectors into matrix $\boldsymbol{V}$ is called the Vandermonde basis $\boldsymbol{V}$ of the null space.

Notice that in the multivariate setting, it is necessary to carefully order the monomials, for which we have chosen to use the degree negative lexicographic ordering, but the method can be easily generalized to any (graded) monomial ordering.

**Definition 3** (Degree negative lexicographic order). Let $\boldsymbol{\alpha}, \boldsymbol{\beta} \in \mathbb{N}^n$ be monomial exponent vectors. Then two monomials are ordered $\boldsymbol{z}^{\boldsymbol{\alpha}} < \boldsymbol{z}^{\boldsymbol{\beta}}$ by the degree negative lexicographic order if $|\boldsymbol{\alpha}| < |\boldsymbol{\beta}|$, or $|\boldsymbol{\alpha}| = |\boldsymbol{\beta}|$ and in the vector difference $\boldsymbol{\beta} - \boldsymbol{\alpha} \in \mathbb{Z}^n$, the left-most non-zero entry is negative.

**Example 4.** The monomials of maximal total degree three in two variables are ordered by the degree negative lexicographic order as

$$1 < z_1 < z_2 < z_1^2 < z_1 z_2 < z_2^2 < z_1^3 < z_1^2 z_2 < z_1 z_2^2 < z_2^3. \quad (27)$$

$\triangle$

### 3.2 System theoretic interpretation

Polynomials are associated with difference equations through the use of the (multidimensional) shift operator $\boldsymbol{z} = (z_1, \ldots, z_n)$ defined in (17). A system of $n$ multivariate polynomial equations in $n$ equations can thus be associated with a set of $n$ difference equations

$$\boldsymbol{r}(\boldsymbol{z})w = \boldsymbol{0}, \quad (28)$$

where $\boldsymbol{r}(\boldsymbol{z}) \in \mathbb{R}^{n\times 1}$ is a vector with polynomial entries. The Macaulay matrix construction can be interpreted as a way to generate equations that $w$ has to satisfy: the rows of the Macaulay matrix are shifts of the difference equations (28). Components of a vector in the null space of the Macaulay matrix can be in this way be seen as shifted samples of $w$.

## 4 From the shift structure to eigenvalue decompositions

In the current section we will illustrate how the multiplicative shift structure of the null space of the Macaulay matrix will lead to the formulation of a state-space realization of the system. In the language of polynomial system solving, this leads to the formulation of the eigenvalue problems of Stetter (2004). In the systems theory framework, it is the application of the shift trick from Ho and Kalman (1966), which leads to the derivation of a corresponding state-space realization.

### 4.1 The shift trick

Let us study the multiplicative shift structure of the null space. Multiplication by monomial $z_i$ maps all total degree $\delta$ monomials to total degree $\delta + 1$ monomials. In general, this is expressed in the Vandermonde monomial vector $\boldsymbol{v}$ as $\boldsymbol{S}_0 \boldsymbol{v} z_i = \boldsymbol{S}_i \boldsymbol{v}$, where $\boldsymbol{S}_0$ selects all monomial rows of total degrees 0 through $d-1$ and $\boldsymbol{S}_i$ selects the rows onto which they are mapped by multiplication with $z_i$. The shift relation for the entire Vandermonde basis $\boldsymbol{V}$ of the null space is

$$\boldsymbol{S}_0 \boldsymbol{V} \boldsymbol{D}_i = \boldsymbol{S}_i \boldsymbol{V}, \tag{29}$$

where $\boldsymbol{D}_i = \mathrm{diag}\left(z_i^{(1)}, \dots, z_i^{(m)}\right)$ contains on the diagonal the evaluation of the shift monomial $z_i$ at the $m$ roots.

In general, the Vandermonde basis $\boldsymbol{V}$ of the null space cannot be obtained directly, but instead a numerical basis $\boldsymbol{Z}$ can be computed, for instance with the singular value decomposition. Recall that the shift relation (29) holds for any basis $\boldsymbol{Z}$ of the null space, which leads to the affine root-finding procedure.

**Theorem 1** (Root-finding (affine)). *Let $\boldsymbol{Z}$ be a basis of the null space of $\boldsymbol{M}$, which is related to the Vandermonde basis by $\boldsymbol{V} = \boldsymbol{Z}\boldsymbol{T}$. The shift relation (29) reduces to the generalized eigenvalue problem*

$$\boldsymbol{S}_0 \boldsymbol{Z} \boldsymbol{T} \boldsymbol{D}_i \boldsymbol{T}^{-1} = \boldsymbol{S}_i \boldsymbol{Z}, \tag{30}$$

*where $\boldsymbol{S}_0$ selects the rows of $\boldsymbol{Z}$ that correspond to the monomials of total degrees 0 through $d-1$, and $\boldsymbol{S}_i$ selects the rows onto which these monomials are mapped under multiplication by $z_i$. The eigenvalues (i.e., the diagonal elements of $\boldsymbol{D}_i$) correspond to the $z_i$ components of the solutions of (18).*

Remark that $\boldsymbol{S}_0 \boldsymbol{Z}$ needs to have full column rank in order to ensure that the eigenvalue problem is not degenerate (i.e., it does not have infinite eigenvalues). In general $\boldsymbol{S}_0 \boldsymbol{Z}$ is nonsquare (tall), which leads to a rectangular generalized eigenvalue problem. We can convert it to a square regular eigenvalue problem by means of the pseudoinverse as $(\boldsymbol{S}_0 \boldsymbol{Z})^\dagger \boldsymbol{S}_i \boldsymbol{Z} = \boldsymbol{T} \boldsymbol{D}_i \boldsymbol{T}^{-1}$.

**Corollary 1** (Reconstructing the Vandermonde basis $\boldsymbol{V}$ from $\boldsymbol{Z}$). *The Vandermonde basis of the null space $\boldsymbol{V}$ can be recovered (up to column-wise scaling) from*

$$\boldsymbol{V} = \boldsymbol{Z}\boldsymbol{T}, \tag{31}$$

*in which all solutions $(z_1^{(k)}, \dots, z_n^{(k)})$ can be read off.*

### 4.2 Multidimensional realization

Similar to the one-dimensional case, we are able to associate with the null space of the Macaulay matrix the interpretation of a multidimensional observability matrix. In Theorem 4 we will further elaborate on this fact.

Let us again consider the column echelon basis of the null space, which we will denote by $\boldsymbol{H}$. In $\boldsymbol{H}$, each of the $m$ columns contains as the first nonzero element a "1" in the rows that correspond to linearly independent monomials. More specifically, they are the lowest-degree linearly independent monomials, ordered by the degree negative lexicographic order. It can be verified that the transformation $\boldsymbol{U}$ has a particular structure in this case.

**Proposition 1.** *Let $\boldsymbol{V} = \boldsymbol{H}\boldsymbol{U}$ express the relation between the column echelon basis $\boldsymbol{H}$ of the null space and the Vandermonde basis $\boldsymbol{V}$ of the null space. The k-th column of $\boldsymbol{U}$ is a monomial vector containing the linearly independent monomials, evaluated at the k-th solution $\left(z_1^{(k)}, \ldots, z_n^{(k)}\right)$.*

*Proof.* Let $\boldsymbol{z}^{\boldsymbol{\alpha}_1}, \boldsymbol{z}^{\boldsymbol{\alpha}_2}, \ldots, \boldsymbol{z}^{\boldsymbol{\alpha}_m}$ denote the linearly independent monomials. Then for a single Vandermonde vector $\boldsymbol{v}$ we have $\boldsymbol{v} = \boldsymbol{H}\boldsymbol{u}$ such that

$$\begin{bmatrix} \boldsymbol{z}^{\boldsymbol{\alpha}_1} \\ \vdots \\ \boldsymbol{z}^{\boldsymbol{\alpha}_2} \\ \vdots \\ \boldsymbol{z}^{\boldsymbol{\alpha}_m} \\ \vdots \\ \boldsymbol{z}^{\boldsymbol{\alpha}_{m-1}} \\ \vdots \end{bmatrix} = \begin{bmatrix} \textcircled{1} & 0 & \cdots & 0 & 0 \\ \times & \boldsymbol{0} & \cdots & 0 & 0 \\ 0 & \textcircled{1} & \ddots & \vdots & \vdots \\ \times & \times & \ddots & \boldsymbol{0} & \boldsymbol{0} \\ \vdots & \ddots & \ddots & \textcircled{1} & 0 \\ \times & \cdots & \times & \times & \boldsymbol{0} \\ 0 & \cdots & 0 & 0 & \textcircled{1} \\ \times & \cdots & \times & \times & \times \end{bmatrix} \begin{bmatrix} \boldsymbol{z}^{\boldsymbol{\alpha}_1} \\ \boldsymbol{z}^{\boldsymbol{\alpha}_2} \\ \vdots \\ \boldsymbol{z}^{\boldsymbol{\alpha}_{m-1}} \\ \boldsymbol{z}^{\boldsymbol{\alpha}_m} \end{bmatrix}, \tag{32}$$

where the circled "ones" are in the linearly independent monomial rows. In the affine case we have $\boldsymbol{z}^{\boldsymbol{\alpha}_1} = 1$. □

The column echelon basis $\boldsymbol{H}$ of the null space allows for a natural interpretation in a multidimensional systems setting.

**Theorem 2** (Canonical realization). *The difference equations $\boldsymbol{r}(\boldsymbol{z})w = \boldsymbol{0}$ admit the state-space realization*

$$\begin{aligned} \boldsymbol{x_H}[k_1+1, k_2, \ldots, k_{n-1}, k_n] &= \boldsymbol{A}_1 \boldsymbol{x_H}[k_1, \ldots, k_n], \\ &\vdots \\ \boldsymbol{x_H}[k_1, k_2, \ldots, k_{n-1}, k_n+1] &= \boldsymbol{A}_n \boldsymbol{x_H}[k_1, \ldots, k_n], \\ y[k_1, \ldots, k_n] &= \boldsymbol{c}^\top \boldsymbol{x_H}[k_1, \ldots, k_n], \end{aligned} \tag{33}$$

*with $\boldsymbol{c} \in \mathbb{R}^m$. The matrices $\boldsymbol{A}_i$, for $i = 1, \ldots, n$, are defined as*

$$\boldsymbol{A}_i := (\boldsymbol{S}_0 \boldsymbol{H})^\dagger \boldsymbol{S}_i \boldsymbol{H}, \tag{34}$$

*and $\boldsymbol{H}$ denotes the column echelon basis of the null space, and $\boldsymbol{x_H}[k_1, \ldots, k_n]$ contain the $w[k_1, \ldots, k_n]$ corresponding to the linearly independent monomials (Proposition 1). The row vector $\boldsymbol{c}^T$ is found as the top row of $\boldsymbol{H}$. The initial conditions that are necessary to iterate the state-space realization (33) can be read off immediately from $\boldsymbol{x_H}[0, \ldots, 0]$.*

**Corollary 2.** *For any basis $\boldsymbol{Z}$ of the null space, where $\boldsymbol{Z} = \boldsymbol{H}\boldsymbol{W}$, one can obtain a state-space realization that is equivalent under a linear state transformation.*

*Proof.* Let $\boldsymbol{Z} = \boldsymbol{H}\boldsymbol{W}$ (where $\boldsymbol{W} = \boldsymbol{U}\boldsymbol{T}^{-1}$ in agreement with earlier definitions). Then it can be verified that the relation (34) becomes $\boldsymbol{W}^{-1}\boldsymbol{A}_i\boldsymbol{W}^{-1} = (\boldsymbol{S}_0\boldsymbol{Z})^\dagger \boldsymbol{S}_i \boldsymbol{Z}$. Furthermore, one can easily verify that this corresponds to a linear state transform $\boldsymbol{x}[k_1, \ldots, k_n] = \boldsymbol{W}\tilde{\boldsymbol{x}}[k_1, \ldots, k_n]$, where $\tilde{\boldsymbol{A}}_i = \boldsymbol{W}^{-1}\boldsymbol{A}_i\boldsymbol{W}$ and $\tilde{\boldsymbol{c}}^\top = \boldsymbol{c}^\top \boldsymbol{W}$. □

**Example 5.** We revisit Example 3 and demonstrate the corresponding canonical realization. The column echelon basis $\boldsymbol{H}$ of the null space is computed as

$$\boldsymbol{H} = \begin{bmatrix} 1 & 0 \\ 0 & 1 \\ 7 & -2 \\ -6 & 5 \\ 12 & -3 \\ 25 & -8 \end{bmatrix}, \tag{35}$$

in which the first two rows correspond to the linearly independent monomials 1 and $z_1$. This implies that $\boldsymbol{S}_0$ selects the first two rows, $\boldsymbol{S}_1$ selects rows two and four and $\boldsymbol{S}_2$ selects rows three and five. A canonical state-space realization can be read off as

$$\begin{aligned} \begin{bmatrix} w[k_1+1, k_2] \\ w[k_1+2, k_2] \end{bmatrix} &= \begin{bmatrix} 0 & 1 \\ -6 & 5 \end{bmatrix} \begin{bmatrix} w[k_1, k_2] \\ w[k_1+1, k_2] \end{bmatrix}, \\ \begin{bmatrix} w[k_1, k_2+1] \\ w[k_1+1, k_2+1] \end{bmatrix} &= \begin{bmatrix} 7 & -2 \\ 12 & -3 \end{bmatrix} \begin{bmatrix} w[k_1, k_2] \\ w[k_1+1, k_2] \end{bmatrix}, \\ y[k_1, k_2] &= \begin{bmatrix} 1 & 0 \end{bmatrix} \begin{bmatrix} w[k_1, k_2] \\ w[k_1+1, k_2] \end{bmatrix}. \end{aligned} \tag{36}$$

△

Remark that the equivalence of polynomial system solving and eigenvalue problems is central in the Stetter approach (Auzinger & Stetter, 1988; Stetter, 2004), where the matrices $\boldsymbol{A}_i := (\boldsymbol{S}_0 \boldsymbol{H})^\dagger \boldsymbol{S}_i \boldsymbol{H}$ correspond to the multiplication matrices that represent multiplication by $z_i$ in the quotient space $\mathbb{C}[z_1, \ldots, z_n]/\langle f_1, \ldots, f_n \rangle$. The linearly independent monomials $\boldsymbol{z}^{\alpha_i}$ correspond to the Groebner basis normal set elements, which form a basis of the quotient space, as discussed in (Batselier, Dreesen, & De Moor, 2014a; Dreesen, 2013). In Batselier and Wong (2016), the following generalization of the Cayley-Hamilton theorem is proved for overdetermined systems.

**Theorem 3.** *(Multidimensional Cayley-Hamilton theorem (affine)) For a set of generators $\{f_1, \ldots, f_n\}$ of the ideal of state difference polynomials $\langle f_1, \ldots, f_n \rangle$ we have that $p(A_1, \ldots, A_n) = 0$ for all $p \in \langle f_1, \ldots, f_n \rangle$.*

The Cayley Hamilton theorem hence implies that the shift matrices $\boldsymbol{A}_1, \ldots, \boldsymbol{A}_n$ satisfy all polynomials that lie in the ideal spanned by the difference equations. As a consequence, we can show that the null space of the Macaulay matrix is a multidimensional generalization of the observability matrix. We first define this particular multidimensional observability matrix.

**Definition 4.** Let $\{f_1, \ldots, f_n\}$ be a set of difference polynomials and $\boldsymbol{A}_1, \ldots, \boldsymbol{A}_n, \boldsymbol{c}^\top$ the corresponding state space parameters. Then we define the corresponding multidimensional extended observability matrix $\mathcal{O}_d$ of total degree $d$ as the matrix

$$\mathcal{O}_d := \begin{bmatrix} \boldsymbol{c}^\top \\ \boldsymbol{c}^\top \boldsymbol{A}_1 \\ \vdots \\ \boldsymbol{c}^\top \boldsymbol{A}_n \\ \boldsymbol{c}^\top \boldsymbol{A}_1^2 \\ \boldsymbol{c}^\top \boldsymbol{A}_1 \boldsymbol{A}_2 \\ \vdots \\ \boldsymbol{c}^\top \boldsymbol{A}_n^2 \\ \vdots \\ \boldsymbol{c}^\top \boldsymbol{A}_1^d \\ \vdots \\ \boldsymbol{c}^\top \boldsymbol{A}_n^d \end{bmatrix}.$$

**Theorem 4.** Let $\{f_1, \ldots, f_n\}$ be a set of difference polynomials and $\boldsymbol{A}_1, \ldots, \boldsymbol{A}_n, \boldsymbol{c}^\top$ the corresponding state space parameters. The null space of the Macaulay matrix $\boldsymbol{M}_d$ is the multidimensional extended observability matrix $\boldsymbol{\mathcal{O}}_d$.

*Proof.* Any element of the row space of the Macaulay matrix $\boldsymbol{M}_d$ contains the coefficients of a polynomial $p \in \langle f_1, \ldots, f_n \rangle$. Multiplying the coefficient vector $\boldsymbol{p}$ with the observability matrix $\boldsymbol{\mathcal{O}}_d$ can be rewritten as $\boldsymbol{c}^T p(\boldsymbol{A}_1, \ldots, \boldsymbol{A}_n)$, which vanishes for any $p \in \langle f_1, \ldots, f_n \rangle$ due to the Cayley-Hamilton theorem. □

**Example 6.** We continue Examples 3 and 5 and use the canonical realization matrices to demonstrate that the observability matrix is the null space of the Macaulay matrix. We have that

$$\begin{aligned}
f_1(\boldsymbol{A}_1, \boldsymbol{A}_2) &= 4\boldsymbol{A}_1^2 - 16\boldsymbol{A}_1 + \boldsymbol{A}_2^2 - 2\boldsymbol{A}_2 + 13\boldsymbol{I}, \\
&= \begin{bmatrix} -24 & 20 \\ -120 & 76 \end{bmatrix} + \begin{bmatrix} 0 & -16 \\ 96 & -80 \end{bmatrix} + \begin{bmatrix} 25 & -8 \\ 48 & -15 \end{bmatrix} + \begin{bmatrix} -14 & 4 \\ -24 & 6 \end{bmatrix} + \begin{bmatrix} 13 & 0 \\ 0 & 13 \end{bmatrix} \\
&= \begin{bmatrix} 0 & 0 \\ 0 & 0 \end{bmatrix},
\end{aligned} \qquad (37)$$

$$\begin{aligned}
f_2(\boldsymbol{A}_1, \boldsymbol{A}_2) &= 2\boldsymbol{A}_1 + \boldsymbol{A}_2 - 7\boldsymbol{I}, \\
&= \begin{bmatrix} 0 & 2 \\ -12 & 10 \end{bmatrix} + \begin{bmatrix} 7 & -2 \\ 12 & -3 \end{bmatrix} + \begin{bmatrix} -7 & 0 \\ 0 & -7 \end{bmatrix} \\
&= \begin{bmatrix} 0 & 0 \\ 0 & 0 \end{bmatrix}.
\end{aligned}$$

Hence, any linear combination $\boldsymbol{c}^T$ of the rows of $\boldsymbol{M}_d$ will also vanish on the observability matrix. △

### Discussion of the realization procedure

Let us review the key properties of the null space of the Macaulay matrix $\boldsymbol{M}$. First, for any choice of basis of the null space, we can use the shift structure to set up an eigenvalue problem that reveals the common roots. Special choices are the Vandermonde basis $\boldsymbol{V}$, in which case the eigenvectors are the columns of the identity matrix, and the column echelon basis $\boldsymbol{H}$, in which case the shift matrix is in Frobenius companion form. The 'canonical' form that is encountered when the column echelon basis $\boldsymbol{H}$ is used, leads to a state-space realization in which the state variables can be interpreted as outputs and their shifts. Remark that the column echelon basis will not be used in practice; instead, one employs a numerical basis $\boldsymbol{Z}$ to compute the roots, which can reliably be computed using the singular value decomposition. Another peculiar property is that the indices of the linearly independent rows, starting from the top, are the same for any choice of basis in the null space of the Macaulay matrix. Because of the latter property, a generalized square eigenvalue problem can be obtained by letting $\boldsymbol{S}_0$ select the $m$ first linearly independent rows of the null space.

### 4.3 Multiple roots

Let us briefly discuss the case of multiple roots. For a system of multivariate polynomials, a $\mu$-fold solution $\boldsymbol{z}^\star$ gives rise to a null space spanned by linear combinations of vectors of the form

$$\frac{1}{\alpha_1! \cdots \alpha_n!} \left. \frac{\partial^{\alpha_1 + \cdots + \alpha_n} \boldsymbol{v}}{\partial^{\alpha_1} z_1 \cdots \partial^{\alpha_n} z_n} \right|_{\boldsymbol{z}^\star}, \qquad (38)$$

where the factor $(\alpha_1! \cdots \alpha_n!)^{-1}$ serves as a normalization. For a thorough treatment of the so-called dual space of $\boldsymbol{M}$, we refer to (Batselier, Dreesen, & De Moor, 2014b; Dayton, Li, & Zeng, 2011). The shift relation in the Vandermonde basis involves in the case of multiple roots a Jordan-like normal form

$$\boldsymbol{S}_0 \boldsymbol{V} \boldsymbol{J}_i = \boldsymbol{S}_i \boldsymbol{V}, \tag{39}$$

where $\boldsymbol{J}_i$ is uppertriangular with $z_i$, evaluated at the $m$ roots, on the diagonal. Some uppertriangular elements of $\boldsymbol{J}_i$ are nonzero, which can be analyzed by inspection of $\boldsymbol{V}$. In the same way as for the one-dimensional case, the occurrence of a Jordan normal form gives rise to so-called Jordan chains in the state-space realization. In practice, the computation of the Jordan normal form is numerically ill-posed, and can be avoided by computing a Schur decomposition (Batselier et al., 2014b).

Let us illustrate the multiplicity structure of the null space for a system of polynomial equations having a four-fold root.

**Example 7.** Consider the equations

$$\begin{array}{rcl} f_1(z_1, z_2) & = & (z_2 - 2)^2 & = & 0, \\ f_2(z_1, z_2) & = & (z_1 - z_2 + 1)^2 & = & 0, \end{array}$$

having a four-fold solution $(1, 2)$. It can be verified that $\boldsymbol{M}$ has a four-dimensional null space that is spanned by the vectors $\partial_{00}$, $\partial_{10}$, $\partial_{01}$ and $2\partial_{20} + \partial_{11}$, where we use a simplified notation $\partial_{\alpha_1 \cdots \alpha_n}$ for (38). We have

$$\boldsymbol{V} = \begin{bmatrix} 1 & 0 & 0 & 0 \\ z_1 & 1 & 0 & 0 \\ z_2 & 0 & 1 & 0 \\ z_1^2 & 2z_1 & 0 & 2 \\ z_1 z_2 & z_2 & z_1 & 1 \\ z_2^2 & 0 & 2z_2 & 0 \\ z_1^3 & 3z_1^2 & 0 & 6z_1 \\ z_1^2 z_2 & 2z_1 z_2 & z_1^2 & 2z_2 + 2z_1 \\ z_1 z_2^2 & z_2^2 & 2z_1 z_2 & 2z_2 \\ z_2^3 & 0 & 3z_2^2 & 0 \end{bmatrix},$$

with $z_1 = 1$ and $z_2 = 2$. It can be verified that this leads to generalized shift relations $\boldsymbol{S}_0 \boldsymbol{V} \boldsymbol{J}_i = \boldsymbol{S}_i \boldsymbol{V}$, for $i = 1, 2$, where

$$\boldsymbol{J}_1 = \begin{bmatrix} 1 & 1 & 0 & 0 \\ 0 & 1 & 0 & 2 \\ 0 & 0 & 1 & 1 \\ 0 & 0 & 0 & 1 \end{bmatrix}, \quad \text{and} \quad \boldsymbol{J}_2 = \begin{bmatrix} 2 & 0 & 1 & 0 \\ 0 & 2 & 0 & 1 \\ 0 & 0 & 2 & 0 \\ 0 & 0 & 0 & 2 \end{bmatrix}. \tag{40}$$

△

## 5 Projective case and descriptor systems

In this section we will shift gears and take a closer look at, and refine where necessary, some of the results that we have obtained so far. We will study roots at infinity, which are 'special' solutions that are caused by algebraic relations among the coefficients. The system theoretic interpretation leads to multidimensional descriptor systems.

### 5.1 Roots at infinity

Solutions at infinity are caused by algebraic relations among the coefficients (often due to the occurrence of zero coefficients) in the equations (18). The Bezout number $m = \prod_i d_i$ counts both affine solutions and solutions at infinity, including multiplicities. Roots at infinity can be analyzed by embedding the system into the $(n + 1)$-dimensional projective space. A homogenization variable $z_0$ is introduced to lift each of the terms of equation $f_i$ to the same total degree $d_i$, denoted by $f_i^h$. The system of homogeneous equations $f_1^h(z_0, z_1, \ldots, z_n) = \cdots = f_n^h(z_0, z_1, \ldots, z_n) = 0$ describes a projective variety, where for affine roots $z_0 = 1$, while for roots at infinity $z_0 = 0$.

**Example 8.** Consider the equations

$$\begin{aligned} f_1(z_1, z_2) &= z_2 - z_1^2 = 0, \\ f_2(z_1, z_2) &= z_1 - 3 = 0, \end{aligned} \tag{41}$$

which has a single affine solution $(3, 9)$. Notice that the Bezout number is $m = 2$, which indicates that the system has two solutions in the projective case. Indeed, by homogenizing the system we have

$$\begin{aligned} f_1^h(z_0, z_1, z_2) &= z_0 z_2 - z_1^2 = 0, \\ f_2^h(z_0, z_1, z_2) &= z_1 - 3z_0 = 0, \end{aligned} \tag{42}$$

where the homogenization variable $z_0$ is introduced. This system has two solutions: an affine solution $(1, 3, 9)$ and a solution at infinity $(0, 0, 1)$. △

The following proposition is essential to understand that roots at infinity are naturally showing up in the Macaulay construction.

**Proposition 2.** *The homogeneous Macaulay coefficient matrix, built from the homogeneous system $f_1^h = \cdots = f_n^h = 0$ is identical to the Macaulay matrix $\boldsymbol{M}$ in Definition 2.*

This fact can be understood by considering the monomials in $n = 2$ variables of total degree $d = 3$ for both the nonhomogeneous and the homogeneous case. We have

$$\left\{ 1, z_1, z_2, z_1^2, z_1 z_2, z_2^2, z_1^3, z_1^2 z_2, z_1 z_2^2, z_2^3 \right\},$$

and

$$\left\{ z_0^3, z_0^2 z_1, z_0^2 z_2, z_0 z_1^2, z_0 z_1 z_2, z_0 z_2^2, z_1^3, z_1^2 z_2, z_1 z_2^2, z_2^3 \right\},$$

both of which are ordered by the degree negative lexicographic ordering. Notice that they are element-wise equal after a substitution $z_0 = 1$.

Due to the property that $z_0 = 0$, roots at infinity give rise to linearly independent monomials of (affine) total degree $d$. This can appreciated from the fact that a substitution $z_0 = 0$ in the homogenized system eliminates all but the terms of total degree $d$. Keeping in mind that all variables $z_0, z_1, \ldots, z_n$ in the projective space are treated on equal footing, it is more suitable to partition roots into regular and singular, denoted by subscripts "$R$" and "$S$", respectively. For affine root finding, the regular roots are the affine roots, while the singular roots are the roots at infinity. As $d$ grows, the linearly independent monomials corresponding to the roots at infinity will therefore move to higher degrees, whereas the linearly independent monomials corresponding to the affine roots will stabilize at low degrees. A simple mechanism to separate affine roots and roots at infinity is hence investigating the row indices of the linearly independent monomials as $d$ increases. The $m_R$ affine linearly independent monomials are of total degrees $\delta \leq d_R$ and stabilize as $d$ increases. The $m_S$ roots at infinity give rise to linearly independent monomials of total degrees $\delta \geq d_S$ with $d_S > d_R$ that move along to higher degrees as $d$ increases.

**Proposition 3** (Mind-the-gap (Dreesen, 2013))**.** *Let the Bezout number $m = m_R + m_S$, where $m_R$ denotes the number of affine roots, and $m_S$ denotes the number of roots at infinity, both of which counted with multiplicity. Furthermore, $d_R$ and $d_S$ are the total degrees defined as above. The total degree at which $d_S > d_R$, i.e., an (affine) degree block is in between the linearly independent monomials corresponding to the affine roots and projective roots, is called the degree of regularity $d^\star$.*

Remark that linear dependence of monomials (rows of the basis of the null space) need not be checked row by row in order to determine $d^\star$. It suffices to monitor for a given total degree $d$ the increase in (numerical) rank of the (affine) total degree $\delta$ blocks of the basis $\boldsymbol{Z}$ of the null space, for $\delta = 0, \ldots, d$. As soon as the rank does not increase in consecutive degree blocks, we have $d^\star = d$.

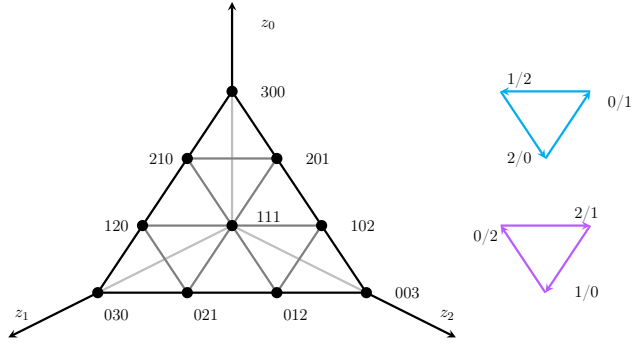

Figure 1: Shifts in the three-dimensional monomial grid take place in the grid points that have an equal total degree. The black dots represent the homogeneous monomials of total degree three. The label $abc$ represents monomial $z_0^a z_1^b z_2^c$. The "$i/j$" shift relation in the homogeneous case amounts to increasing the exponent of $z_i$ by one, and decreasing the exponent of $z_j$ by one. In the monomial grid a shift is a move from a point to an adjacent point. The six possible shifts, i.e., 0/1, 0/2, 1/0, 1/2, 2/0 and 2/1, are denoted by the arrows. Notice that not all shifts are possible: monomials along the edges have exponents equal to zero that cannot be decreased.

### 5.2 Projective shift relation

Let us investigate how the shift relation can be generalized to the homogeneous case. The exposition is similar to that of Batselier et al. (2014b), but we review here the main facts for completeness.

**Proposition 4** (Homogeneous shift relation). *Let $\boldsymbol{V}$ denote the homogeneous Vandermonde basis of the null space. We can write the shift relation as*

$$\boldsymbol{S}_{i/j}\boldsymbol{V}\boldsymbol{D}_i = \boldsymbol{S}_{j/i}\boldsymbol{V}\boldsymbol{D}_j, \tag{43}$$

*with row selection matrices $\boldsymbol{S}_{i/j}$ and $\boldsymbol{S}_{j/i}$ describing an up shift in $z_i$ and a down shift in $z_j$, and $\boldsymbol{D}_i$ and $\boldsymbol{D}_j$ are diagonal matrices with the values of $z_i$ and $z_j$ on the diagonals.*

Figure 1 is a visual representation of the shift relation for the $n = 3$ and $d = 3$ case.

For a numerical basis $\boldsymbol{Z}$ of the null space with $\boldsymbol{V} = \boldsymbol{Z}\boldsymbol{T}$ this leads to the homogeneous eigenvalue problem

$$\boldsymbol{S}_{i/j}\boldsymbol{Z}\boldsymbol{T}\boldsymbol{D}_i = \boldsymbol{S}_{j/i}\boldsymbol{Z}\boldsymbol{T}\boldsymbol{D}_j. \tag{44}$$

**Example 9.** The "1/2" shift relation of Proposition 4 for the case $n = 2$ and $d = 2$ is

$$\begin{bmatrix} 0 & 0 & 1 & 0 & 0 & 0 \\ 0 & 0 & 0 & 0 & 1 & 0 \\ 0 & 0 & 0 & 0 & 0 & 1 \end{bmatrix} \begin{bmatrix} z_0^2 z_1^0 z_2^0 \\ z_0^1 z_1^1 z_2^0 \\ z_0^1 z_1^0 z_2^1 \\ z_0^0 z_1^2 z_2^0 \\ z_0^0 z_1^1 z_2^1 \\ z_0^0 z_1^0 z_2^2 \end{bmatrix} z_1$$
$$= \begin{bmatrix} 0 & 1 & 0 & 0 & 0 & 0 \\ 0 & 0 & 0 & 1 & 0 & 0 \\ 0 & 0 & 0 & 0 & 1 & 0 \end{bmatrix} \begin{bmatrix} z_0^2 z_1^0 z_2^0 \\ z_0^1 z_1^1 z_2^0 \\ z_0^1 z_1^0 z_2^1 \\ z_0^0 z_1^2 z_2^0 \\ z_0^0 z_1^1 z_2^1 \\ z_0^0 z_1^0 z_2^2 \end{bmatrix} z_2. \tag{45}$$

△

As long as either $z_i \neq 0$ or $z_j \neq 0$ for all roots, the homogeneous eigenvalue problem (44) can always be reduced to an affine eigenvalue problem. If $z_j \neq 0$ we can write

$$\boldsymbol{S}_{i/j}\boldsymbol{ZTD}_{i/j}\boldsymbol{T}^{-1} = \boldsymbol{S}_{j/i}\boldsymbol{Z}, \tag{46}$$

with $\boldsymbol{D}_{i/j} = \boldsymbol{D}_i\boldsymbol{D}_j^{-1}$. However, if $z_j = 0$ then inverting $\boldsymbol{D}_j$ is not possible.

Remark that, in the case of affine root-finding (Theorem 1), we consider shifts up in the affine components $z_1, z_2, \ldots, z_n$, whereas roots at infinity are characterized by $z_0 = 0$, which is the degenerate component that is being shifted down (implicitly).

In agreement with the separation of the linearly independent monomials into regular and singular roots, the column echelon basis $\boldsymbol{H}$ of the null space can be partitioned accordingly. We have now

$$\boldsymbol{H} = \begin{bmatrix} \boldsymbol{H}_R^{(1)} & \boldsymbol{0} \\ \boldsymbol{H}_R^{(2)} & \boldsymbol{0} \\ \times & \boldsymbol{H}_S \end{bmatrix}, \tag{47}$$

where $\boldsymbol{H}_R^{(1)}$ and $\boldsymbol{H}_S$ denote the (affine) degree blocks containing the regular and singular linearly independent monomials, respectively, and $\boldsymbol{H}_R^{(2)}$ denotes the (affine) degree block where there are no linearly independent monomials (but zeros next to it).

For a numerical basis of the null space $\boldsymbol{Z}$, the regular and singular columns are not separated, as every column is in general composed as a linear combination of null space vectors that come from both the regular and singular parts. We perform a column compression to find $\boldsymbol{Z}_R$ having $m_R$ columns and rank $m_R$ on which the affine root-finding procedure can be applied.

**Lemma 3** (Column compression). *Let $\boldsymbol{W} = \begin{bmatrix} \boldsymbol{W}_1^\top & \boldsymbol{W}_2^\top \end{bmatrix}^\top$ be a $q \times m$ matrix, which is partitioned into a $k \times m$ matrix $\boldsymbol{W}_1$ and an $(q-k) \times m$ matrix $\boldsymbol{W}_2$ with $\mathrm{rank}(\boldsymbol{Z}_1) = m_R < m$. Let the singular value decomposition of $\boldsymbol{W} = \boldsymbol{U\Sigma V}^\top$. Then $\boldsymbol{Z} = \boldsymbol{WQ}$ is called the column compression of $\boldsymbol{W}$ and can be partitioned as*

$$\boldsymbol{Z} = \begin{bmatrix} \boldsymbol{Z}_{11} & \boldsymbol{0} \\ \boldsymbol{Z}_{21} & \boldsymbol{Z}_{22} \end{bmatrix}, \tag{48}$$

*where $\boldsymbol{Z}_{11}$ has size $k \times m_R$ (and the remaining blocks have compatible dimensions).*

Once the column compression is obtained, the affine root-finding procedure can be applied in a straightforward way.

**Corollary 3** (Root-finding in the presence of singular roots). *Consider a column compression of the degree-blocks of $\boldsymbol{Z}$ corresponding to the $\boldsymbol{H}_R^{(1)}$ and $\boldsymbol{H}_R^{(2)}$, where*

$$\boldsymbol{Z} = \begin{bmatrix} \boldsymbol{Z}_R^{(1)} & \boldsymbol{0} \\ \boldsymbol{Z}_R^{(2)} & \boldsymbol{0} \\ \times & \times \end{bmatrix}, \tag{49}$$

*where $\boldsymbol{Z}_R$ has $m_R$ columns and is compatible with the shift relation involving the affine linearly independent monomial rows. Applying the affine root-finding procedure of Theorem 1 on $\boldsymbol{Z}_R$ returns the regular roots.*

Remark that the column compression 'from the right' requires a rank decision: This rank, as revealed by the column compression of the upper part of the null space, is equal to the number of affine roots.

### 5.3 Descriptor systems

Affine roots and roots at infinity have an elegant interpretation in terms of (multidimensional) systems. Let us first recall that a pencil $\lambda\boldsymbol{E} - \boldsymbol{A}$ can be realized as a descriptor system.

**Lemma 4** (Weierstrass canonical form). *A pencil $\lambda E - A$ with $A, E \in \mathbb{R}^{m \times m}$ can be decomposed into a regular and a singular part, denoted by $A_R$ and $E_S$, respectively, by the Weierstrass canonical form (Gantmacher, 1960; Gerdin, 2004; Kailath, 1980; Luenberger, 1978)*

$$P(\lambda E - A) Q = \begin{bmatrix} A_R - \lambda I & 0 \\ 0 & \lambda E_S - I \end{bmatrix}, \tag{50}$$

*for some $P, Q \in \mathbb{R}^{m \times m}$. The Weierstrass canonical form admits the following state-space realization (Moonen, De Moor, Ramos, & Tan, 1992)*

$$\begin{bmatrix} x_R[k+1] \\ x_S[k-1] \end{bmatrix} = \begin{bmatrix} A_R & 0 \\ 0 & E_S \end{bmatrix} \begin{bmatrix} x_R[k] \\ x_S[k] \end{bmatrix}, \tag{51}$$

*where $E_S$ is nilpotent, and $x_R$ and $x_S$ denote the regular and singular parts of the state.*

An iteration "running forward in time" is obtained for the regular part of the state, while an iteration "running backward in time" is obtained for the singular part. Notice that the forward and backward iterations are nothing more than the up and down shifts that we encountered before. The separation of regular and singular parts can be interpreted in the polynomial system solving framework as the separation of the affine roots and roots at infinity.

Recall that the projective description allows for an equal treatment of all variables, and can also determine the roots at infinity. In order to include the roots at infinity, one chooses an up shift in $x_0$, implying that the one of the affine components $z_1, \ldots, z_n$ is shifted down.

**Example 10.** Consider the system

$$\begin{array}{rcl} f_1(z_1, z_2) & = & z_1^2 + z_1 z_2 - 10 z_0^2 = 0, \\ f_2(z_1, z_2) & = & z_2^2 + z_1 z_2 - 15 z_0^2 = 0, \end{array} \tag{52}$$

having two affine solutions $(1, 2, 3)$, $(1, -2, -3)$, and a projective root $(0, 1, -1)$ with multiplicity two. At total degree $d = 4$ the linearly independent monomials are $1, z_1, z_1^3$ and $z_1^4$, of which $1$ and $z_1$ are associated with to the two affine roots and $z_1^3$ and $z_1^4$ with the double root at infinity. Notice that there are no linearly independent monomials of total degree two, which means that there is a separation of one degree block between the two sets of linearly independent monomials. The column echelon basis $H$ of the null space is partitioned as in (47), and we find

$$\left[ \begin{array}{c} H_R^{(1)} \\ \hline H_R^{(2)} \end{array} \right] = \begin{bmatrix} 1 & 0 \\ 0 & 1 \\ 0 & 3/2 \\ 4 & 0 \\ 6 & 0 \\ 9 & 0 \end{bmatrix}, \text{ and } \begin{bmatrix} H_S \end{bmatrix} = \begin{bmatrix} 1 & 0 \\ -1 & 0 \\ 1 & 0 \\ -1 & 0 \\ 0 & 1 \\ 0 & -1 \\ 0 & 1 \\ 0 & -1 \\ 0 & 1 \end{bmatrix}. \tag{53}$$

For the affine roots, we can immediately write the corresponding realization problems

$$\begin{bmatrix} w[k_1+1, k_2] \\ w[k_1+2, k_2] \end{bmatrix} = \begin{bmatrix} 0 & 1 \\ 4 & 0 \end{bmatrix} \begin{bmatrix} w[k_1, k_2] \\ w[k_1+1, k_2] \end{bmatrix}$$

$$\begin{bmatrix} w[k_1, k_2+1] \\ w[k_1+1, k_2+1] \end{bmatrix} = \begin{bmatrix} 0 & 3/2 \\ 6 & 0 \end{bmatrix} \begin{bmatrix} w[k_1, k_2] \\ w[k_1+1, k_2] \end{bmatrix} \tag{54}$$

$$y[k_1, k_2] = \begin{bmatrix} 1 & 0 \end{bmatrix} \begin{bmatrix} w[k_1, k_2] \\ w[k_1+1, k_2] \end{bmatrix}$$

where the action matrices $\boldsymbol{A}_1$ and $\boldsymbol{A}_2$ have the eigenvalue decompositions

$$\begin{bmatrix} 0 & 1 \\ 4 & 0 \end{bmatrix} = \begin{bmatrix} 1 & 1 \\ 2 & -2 \end{bmatrix} \begin{bmatrix} 2 & 0 \\ 0 & -2 \end{bmatrix} \begin{bmatrix} 1 & 1 \\ 2 & -2 \end{bmatrix}^{-1},$$

$$\begin{bmatrix} 0 & 3/2 \\ 6 & 0 \end{bmatrix} = \begin{bmatrix} 1 & 1 \\ 2 & -2 \end{bmatrix} \begin{bmatrix} 3 & 0 \\ 0 & -3 \end{bmatrix} \begin{bmatrix} 1 & 1 \\ 2 & -2 \end{bmatrix}^{-1}. \quad (55)$$

We recognize in (53) in the $\boldsymbol{H}_S$ block the evaluation of $(0, 1, -1)$ in the (homogeneous) Vandermonde vector $\boldsymbol{v}$, which has only nonzero elements in the highest (affine) degree block, and the evaluation of $(0, 1, -1)$ in the differential $\partial \boldsymbol{v}/\partial z_0$, which is nonzero only in the $d-1$ degree block. The multiplicity structure of the root $(0, 1, -1)$ can be read off immediately from the relations

$$\boldsymbol{S}_{0/1} \begin{bmatrix} \boldsymbol{v} & \partial \boldsymbol{v}/\partial z_0 \end{bmatrix} \begin{bmatrix} 0 & 1 \\ 0 & 0 \end{bmatrix} = \boldsymbol{S}_{1/0} \begin{bmatrix} \boldsymbol{v} & \partial \boldsymbol{v}/\partial z_0 \end{bmatrix},$$

$$\boldsymbol{S}_{0/2} \begin{bmatrix} \boldsymbol{v} & \partial \boldsymbol{v}/\partial z_0 \end{bmatrix} \begin{bmatrix} 0 & -1 \\ 0 & 0 \end{bmatrix} = \boldsymbol{S}_{2/0} \begin{bmatrix} \boldsymbol{v} & \partial \boldsymbol{v}/\partial z_0 \end{bmatrix}, \quad (56)$$

△

Let us finally remark that in the homogeneous setting it is valid to consider the "$0/i$" shift, where $z_0$ is shifted upwards and $z_i$ is shifted downwards. Since $z_0$ is the homogenization variable, the eigenvalue decomposition $\boldsymbol{S}_{0/i} \boldsymbol{Z} \boldsymbol{D}_{0/i} = \boldsymbol{S}_{i/0} \boldsymbol{V}$ has as many nonzero eigenvalues as there are affine roots, and as many zero eigenvalues as there are roots at infinity. Also in this case, the presented methods allow for computing the solutions.

The separation of the state variables into regular and singular parts allows for a general descriptor system realization as in Batselier and Wong (2016): An autonomous overdetermined multidimensional descriptor system is described over the $n+1$-dimensional state-vector $\boldsymbol{x}$ with an additional equation of the form

$$\boldsymbol{x}[k_0 + 1, k_1, \ldots, k_n] = \boldsymbol{A}_0 \boldsymbol{x}[k_0, \ldots, k_n]. \quad (57)$$

In accordance with the separation of the regular and singular parts, it can be shown (Batselier & Wong, 2016) that the matrices $\boldsymbol{A}_i$ can be partitioned as

$$\boldsymbol{A}_0 = \begin{bmatrix} \boldsymbol{I} & \boldsymbol{0} \\ \boldsymbol{0} & \boldsymbol{E}_0 \end{bmatrix}, \text{ and } \boldsymbol{A}_i = \begin{bmatrix} \boldsymbol{R}_i & \boldsymbol{0} \\ \boldsymbol{0} & \boldsymbol{E}_i \end{bmatrix}, \text{ for } i = 1, \ldots, n, \quad (58)$$

where $\boldsymbol{I} \in \mathbb{R}^{m_R \times m_R}$ denotes the identity matrix, $\boldsymbol{E}_0 \in \mathbb{R}^{m_S \times m_S}$ is a nilpotent matrix that agrees with the roots at infinity; the matrices $\boldsymbol{A}_i$ are partitioned in the same way, i.e., $\boldsymbol{R}_i \in \mathbb{R}^{m_R \times m_R}$ and $\boldsymbol{E}_i \in \mathbb{R}^{m_S \times m_S}$. The separation of the affine roots and the roots at infinity agrees with the partitioning in the matrices $\boldsymbol{A}_i$ and allows for a separation of the state vector $\boldsymbol{x}$ into a regular part $\boldsymbol{x}_R$ and a singular part $\boldsymbol{x}_S$.

## 6 Conclusions

We have shown that simple linear algebra tools allow for the construction of an eigenvalue-based approach for solving systems of polynomial equations, without resorting to computer algebra methods. The system theoretical interpretation was the state space realization problem from a given set of difference equations. Solutions at infinity were described using descriptor systems.

Future work is concerned with involving external input signals as in Ball and Vinnikov (2003); Batselier and Wong (2016); Fornasini et al. (1993), as well as exploring higher-dimensional solution sets and their system theoretic interpretation. Finally, the insights developed in the current article may inspire the development of subspace-based system identification methods (Van Overschee & De Moor, 1996) for certain classes of multidimensional systems.

# Acknowledgments

PD is a postdoctoral researcher at Vrije Universiteit Brussel, Dept. ELEC and a free researcher at KU Leuven, Dept. ESAT-STADIUS. KB is a postdoctoral researcher at The University of Hong Kong, Dept. EEE. BDM is a full professor at KU Leuven, Dept. ESAT-STADIUS. This work was supported in part by: Belgian Federal Science Policy Office: IUAP P7/19 (DYSCO, Dynamical systems, control and optimization, 2012-2017); Flemish Government: IWT/FWO PhD grants; KU Leuven Internal Funds C16/15/059, C32/16/013; imec strategic funding 2017; Fund for Scientific Research (FWO-Vlaanderen); the Flemish Government (Methusalem); the Belgian Government through the Inter university Poles of Attraction (IAP VII) Program; ERC advanced grant SNLSID, under contract 320378; FWO grants G028015N and G090117N. Part of this work was done when PD held a PhD grant of Flanders' Agency for Innovation by Science and Technology (IWT Vlaanderen).

The scientific responsibility is assumed by the authors.

# References

Attasi, S. (1976). Modelling and recursive estimation for double indexed sequences. In *System identification: Advances and case studies* (pp. 289–348). New York: Academic Press.

Auzinger, W., & Stetter, H. J. (1988). An elimination algorithm for the computation of all zeros of a system of multivariate polynomial equations. In *Proc. Int. Conf. Num. Math.* (pp. 11–30). Birkhäuser.

Ball, J. A., Boquet, G. M., & Vinnikov, V. (2012). A behavioral interpretation of Livšic systems. *Multidimens. Syst. Signal Process.*, *23*(1), 17–48.

Ball, J. A., & Vinnikov, V. (2003). Overdetermined multidimensional systems: State space and frequency domain methods. In J. Rosenthal & D. S. Gilliam (Eds.), *Mathematical Systems Theory in Biology,Communications, Computation, and Finance* (pp. 63–119). New York: Springer.

Batselier, K., Dreesen, P., & De Moor, B. (2014a). The canonical decomposition of $C_d^n$ and numerical Gröbner and border bases. *SIAM J. Mat. Anal. Appl.*, *35*(4), 1242–1264.

Batselier, K., Dreesen, P., & De Moor, B. (2014b). On the null spaces of the Macaulay matrix. *Lin. Alg. Appl.*, *460*(1), 259–289.

Batselier, K., & Wong, N. (2016). Computing the state difference equations for discrete overdetermined linear systems. *Automatica*, *64*, 254–261.

Bleylevens, I., Peeters, R., & Hanzon, B. (2007). Efficiency improvement in an nD-systems approach to polynomial optimization. *J. Symb. Comput.*, *42*(1–2), 30–53.

Bose, N. K. (1982). *Applied multidimensional systems theory*. Van Nostrand Reinhold.

Bose, N. K. (2007). Two decades (1985-2005) of Gröbner bases in multidimensional systems. In H. A. Park & G. Regensburger (Eds.), *Gröbner bases in control theory and signal processing* (Vol. 3, pp. 1–22). Walter de Gruyter.

Bose, N. K., Buchberger, B., & Guiver, J. P. (2003). *Multidimensional systems theory and applications*. Springer.

Buchberger, B. (2001). Gröbner bases and systems theory. *Multidimens. Syst. Signal Process.*, *12*, 223–251.

Cox, D. A., Little, J. B., & O'Shea, D. (2005). *Using Algebraic Geometry* (Second ed.). New York: Springer-Verlag.

Cox, D. A., Little, J. B., & O'Shea, D. (2007). *Ideals, Varieties and Algorithms* (Third ed.). Springer-Verlag.

Dayton, B. H., Li, T.-Y., & Zeng, Z. (2011). Multiple zeros of nonlinear systems. *Math. Comp.*, *80*, 2143–2168.

Dreesen, P. (2013). *Back to the roots – polynomial system solving using linear algebra* (Unpublished doctoral dissertation). Faculty of Engineering Science, KU Leuven, Leuven, Belgium.

Fornasini, E., & Marchesini, G. (1976). State-space realization theory of two-dimensional filters. *IEEE T. Automat. Contr.*, *21*, 484–492.

Fornasini, E., Rocha, P., & Zampieri, S. (1993). State space realization of 2-D finite-dimensional behaviors. *SIAM J. Control and Optimization*, *31*(6), 1502–1517.

Gałkowski, K. (2001). *State-space realizations of linear 2-D systems with extensions to the general nD (n > 2) case*. Springer.

Gantmacher, F. (1960). *The theory of matrices* (Vol. 2). New York: Chelsea Publishing Company.

Gerdin, M. (2004, May). Computation of a canonical form for linear differential-algebraic equations. In *Proc. Reglermöte 2004*. Göteborg.

Giusti, M., & Schost, E. (1999). Solving some overdetermined polynomial systems. In *Proc. 1999 Int. Symp. Symb. Algebraic Comput. (ISSAC 1999)* (pp. 1–8). ACM.

Hanzon, B., & Hazewinkel, M. (Eds.). (2006a). *Constructive algebra and systems theory*. Royal Netherlands Academy of Arts and Sciences.

Hanzon, B., & Hazewinkel, M. (2006b). An introduction to constructive algebra and systems theory. In B. Hanzon & M. Hazewinkel (Eds.), *Constructive algebra and systems theory* (pp. 2–7). Royal Netherlands Academy of Arts and Sciences.

Ho, B. L., & Kalman, R. E. (1966). Effective construction of linear state-variable models from input/output functions. *Regelungstechnik*, *14*(12), 545–548.

Jónsson, G. F., & Vavasis, S. A. (2004). Accurate solution of polynomial equations using Macaulay resultant matrices. *Math. Comput.*, *74*(249), 221–262.

Kaczorek, T. (1988). The singular general model of 2D systems and its solution. *IEEE T. Automat. Contr.*, *33*, 1060–1061.

Kailath, T. (1980). *Linear systems*. Prentice-Hall International.

Kurek, J. E. (1985). Basic properties of $q$-dimensional linear digital systems. *Int. J. Control*, *42*, 119–128.

Lazard, D. (1983). Groebner bases, Gaussian elimination and resolution of systems of algebraic equations. In J. van Hulzen (Ed.), *Computer algebra* (Vol. 162, pp. 146–156). Springer Berlin / Heidelberg.

Livšic, M. S. (1983). Cayley-Hamilton theorem, vector bundles and divisors of commuting operators. *Integral Equations Operator Theory*, *6*, 250–273.

Livšic, M. S., Kravitsky, N., Markus, A. S., & Vinnikov, V. (1995). Theory of commuting nonselfadjoint operators. In (Vol. 332). Dordrecht, NL: Kluwer Academic Publisher Group.

Luenberger, D. G. (1978). Time-invariant descriptor systems. *Automatica*, *14*, 473–480.

Macaulay, F. S. (1916). *The algebraic theory of modular systems*. Cambridge University Press.

Möller, H. M., & Stetter, H. J. (1995). Multivariate polynomial equations with multiple zeros solved by matrix eigenproblems. *Numer. Math.*, *70*, 311–329.

Moonen, M., De Moor, B., Ramos, J., & Tan, S. (1992). A subspace identification algorithm for descriptor systems. *Syst. Control Lett.*, *19*, 47–52.

Mourrain, B. (1998). Computing the isolated roots by matrix methods. *J. Symb. Comput.*, *26*, 715–738.

Oberst, U. (1990). Multidimensional constant linear systems. *Acta Appl. Math.*, *20*(1), 1–175.

Ramos, J. A., & Mercère, G. (2016). Subspace algorithms for identifying separable-in-denominator 2D systems with deterministic-stochastic inputs. *Int. J. Control*, *89*, 2584–2610.

Rocha, P., & Willems, J. C. (2006). Markov properties for systems described by PDEs and first-order representations. *Systems & Control Letters*, *55*, 538–542.

Roesser, R. (1975). A discrete state-space model for linear image processing. *IEEE T. Automat. Contr.*, *20*, 1–10.

Rogers, E., Gałkowski, K., Paszke, W., Moore, K. L., Bauer, P. H., Hladowski, L., & Dabkowski, P. (2015). Multidimensional control systems: case studies in design and evaluation. *Multidimens. Syst. Signal Process.*, *26*(4), 895–939.

Shafarevich, I. R. (2013). *Basic algebraic geometry I — Varieties in projective space* (third ed.). Springer.

Shaul, L., & Vinnikov, V. (2009, June). State feedback for overdetermined 2D systems: pole placement for bundle maps over an algebraic curve. In *International workshop on multidimensional (nD) systems (nDS 2009)* (p. 1-2).

Stetter, H. J. (2004). *Numerical polynomial algebra*. SIAM.

Van Overschee, P., & De Moor, B. (1996). *Subspace identification for linear systems: theory, implementation, applications*. Dordrecht, Netherlands: Kluwer Academic Publishers.

Willems, J. C. (1986a). From time series to linear system – Part I. Finite dimensional linear time invariant systems. *Automatica*, *22*(5), 561–580.

Willems, J. C. (1986b). From time series to linear system – Part II. Exact modeling. *Automatica*, *22*(6), 675–694.

Willems, J. C. (1987). From time series to linear system – Part III. Approximate modeling. *Automatica*, *23*(1), 87–115.

Xu, L., Fan, H., Lin, Z., & Bose, N. K. (2008). A direct-construction approach to multidimensional realization and LFR uncertainty modeling. *Multidimensional Systems and Signal Processing*, *22*(1–3), 323–359.

Xu, L., Yan, S., Lin, Z., & Matsushita, S. (2012). A new elementary operation approach to multidimensional realization and LFR uncertainty modeling: the MIMO case. *IEEE Transactions on Circuits and Systems*, *59*(3), 638–651.

Zerz, E. (2000). *Topics in multidimensional linear systems theory*. Springer.

Zerz, E. (2008). The discrete multidimensional MPUM. *Multidim. Syst. Sign. Process.*, *19*, 307–321.